\documentclass[manuscript]{acmart}
\AtBeginDocument{%
  \providecommand\BibTeX{{%
    \normalfont B\kern-0.5em{\scshape i\kern-0.25em b}\kern-0.8em\TeX}}}

\setcopyright{acmcopyright}
\copyrightyear{2021}
\acmYear{2021}
\acmDOI{10.1145/1122445.1122456}

\acmConference[MobileHCI '21]{MobileHCI '21: The ACM International Conference on Mobile Human-Computer Interaction}{Sep 27 -- Oct 1, 2021}{Toulouse, France}
\acmBooktitle{MobileHCI '21: The ACM International Conference on Mobile Human-Computer Interaction, Sep 27 -- Oct 1, 2021, Toulouse, France}
\acmPrice{15.00}
\acmISBN{978-1-4503-XXXX-X/18/06}

\begin{document}

\title{Designing Mobile EEG Neurofeedback Games for Children with Autism}

\author{Zhaoyi Yang}
\email{yangzhaoyivip@gmail.com }
\affiliation{%
  \institution{BrainCo Inc.}
  \city{Somerville}
  \state{Massachusetts}
  \country{USA}
}

\author{Pengcheng An}
\email{anpengcheng88@gmail.com}
\affiliation{%
  \institution{Cheriton School of Computer Science, University of Waterloo}
  \city{Waterloo}
  \state{Ontario}
  \country{Canada}
}

\author{Jinchen Yang}
\affiliation{%
  \institution{BrainCo Inc.}
  \city{Somerville}
  \state{Massachusetts}
  \country{USA}
}

\author{Samuel Strojny}
\affiliation{%
  \institution{BrainCo Inc.}
  \city{Somerville}
  \state{Massachusetts}
  \country{USA}
}

\author{Jinchen Yang}
\affiliation{%
  \institution{BrainCo Inc.}
  \city{Somerville}
  \state{Massachusetts}
  \country{USA}
}

\author{Zihui Zhang}
\affiliation{%
  \institution{BrainCo Inc.}
  \city{Somerville}
  \state{Massachusetts}
  \country{USA}
}

\author{Dongsheng Sun}
\affiliation{%
  \institution{BrainCo Inc.}
  \city{Somerville}
  \state{Massachusetts}
  \country{USA}
}

\author{Jian Zhao}
\email{jianzhao@uwaterloo.ca}
\affiliation{%
  \institution{Cheriton School of Computer Science, University of Waterloo}
  \city{Waterloo}
  \state{Ontario}
  \country{Canada}
}

\renewcommand{\shortauthors}{Yang and An, et al.}

\begin{abstract}
Neurofeedback games are an effective and playful approach to enhance certain social and attentional capabilities in children with autism, which are promising to become widely accessible along with the commercialization of mobile EEG modules. However, little industry-based experiences are shared, regarding how to better design neurofeedback games to fine-tune their playability and user experiences for autistic children. In this paper, we review the experiences we gained from industry practice, in which a series of mobile EEG neurofeedback games have been developed for preschool autistic children. We briefly describe our design and development in a one-year collaboration with a special education center involving a group of stakeholders: children with autism and their caregivers and parents. We then summarize four concrete implications we learnt concerning the design of game characters, game narratives, as well as gameplay elements, which aim to support future work in creating better neurofeedback games for preschool children with autism.
\end{abstract}



\begin{teaserfigure}
  \includegraphics[width=\textwidth]{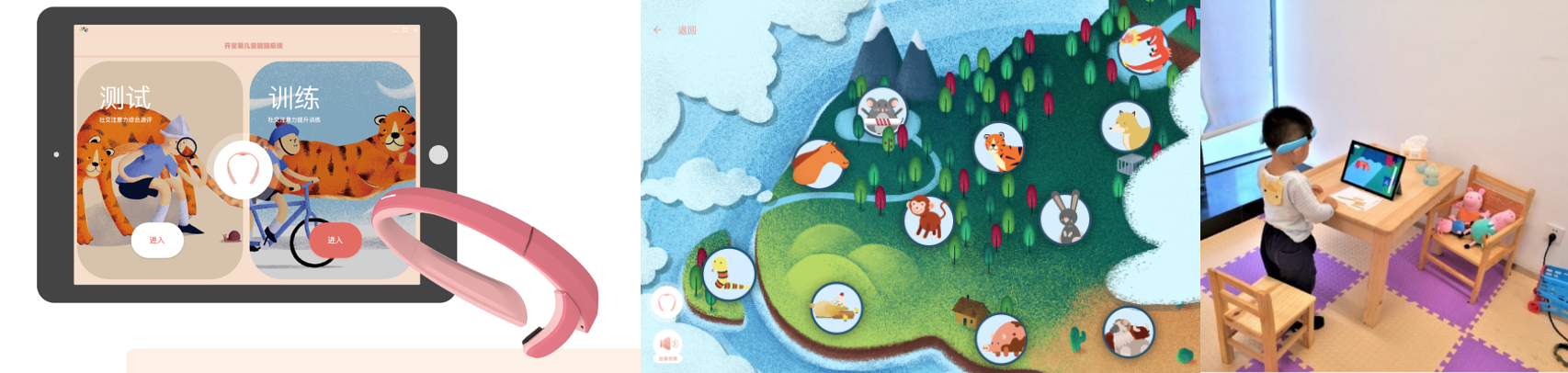}
  \caption{Left and Middle: the neurofeedback game device and game selection screen of the Starkids system (alias). Right: a preschool child with ASD playing with the system in a real-world setting (a special education center in Hangzhou, China)}
  \Description{Left: a tablet and a headset; the tablet displays the starting screen of the Starkids system; Middle: the starting menu of Starkids 12 Zodiac animal games; Right: a preschool child playing the Starkids NFT game in a special education center.}
  \label{fig:teaser}
\end{teaserfigure}

\maketitle

\section{Introduction}
HCI research has seen increased interest in designing interactive systems to support children living with Autism Spectrum Disorder (ASD) \cite{Barakova2009,Frauenberger2016,Marcu2013,Spiel2017}. In daily social situations, children with ASD may face many difficulties due to their impaired social and communication skills (imitation, empathy, joint-attention, etc.), restricted interpersonal interests, as well as repeated behavioral patterns. \cite{Association2013,Schopler1986}. Prior reviews show that therapeutic games can be an appropriate means to help children with ASD in improving social functioning \cite{Friedrich2014}. For instance, games like the Social Mirroring Game \cite{Friedrich2015}, can enable autistic children to watch social interaction sequences, triggering their mirror neuron activity (related to abilities of imitation, intention understanding and empathy \cite{Dapretto2006}). Meanwhile, by giving gamified feedback based on the mu waves \cite{Bernier2007} of their brain signals (an EEG indicator of mirror neuron activity), the game reinforces the children’s social brain functioning \cite{Datko2018,Friedrich2015}.

These interventions are primarily neurofeedback training (NFT), which allows real-time signals of brain activities to be fed back to a user through multimodal representations on a computing interface, thus forming a feedback loop to strengthen users’ self-regulation of certain brain activities \cite{Friedrich2014,Pineda2008}.

Admittedly, the therapeutic benefits of this mirror neuron approach continue to be recognized in neuroscientific or medical studies \cite{Dapretto2006,Datko2018,Friedrich2014,Pineda2008}. However, a critical aspect is how this neurofeedback technique from medical domain should be properly leveraged by HCI design for real-world service-product systems used in daily contexts. Primarily, mirror neuron stimulating is means rather than end: activating mirror neuron does not simply mean positive experience or increased life quality for these users. As HCI research points out \cite{Spiel2017}, adopting merely a medical model of autism may lead to the underlying normality that sees autism only as a problem to be treated, rather than experiences that need to be understood and designed for.

In other word, such neurofeedback interventions should not be designed merely as “medical treatments”, and their experiential and hedonic aspects are as important as their therapeutic benefits. In this work, we thereby aim to preliminarily address the design aspects and implications of NFT games, based on our commercial development and industry practice.

While existing research has examined the effects of NFT games (or interventions) for autistic children \cite{Datko2018,Friedrich2015,Mandryk2013,Mercado2020,Pineda2008}, commercialized applications or services are still in scarcity. Hence practice-based insights in how to design interfaces of such games to optimize their playability and enhance user experiences, is still lacking. To contribute such practical insights, we summarize our experiences gained from the ongoing project of Starkids (alias), a product-service system by BrainCo Inc., targeting at the Chinese market (Figure 1). 

\section{The Starkids system and the game design}
As Figure 1 shows, the Starkids product-service system includes a wearable EEG sensor-kit dedicated for preschool users with ASD (aged 3-6), an application, as well as a series of NFT game contents that are open-ended for continual update and expansion. Starkids service is currently aimed for ad-hoc training settings in which children with ASD use the system with support or guidance from caregivers (professional educators and therapists, see Figure 2), e.g., in a special education center (see Figure 1 Right). Normally, each user will engage in such a training session for 30 minutes, three to five times a week.

The game contents of Starkids currently include 12 animated games and 24 games that use video clips of people in reality. All these games depict occasions involving social interaction and collaboration among children or cartoon characters, which has been proven to be an effective way to stimulate and reinforce social attention of children with ASD \cite{Dapretto2006,Pineda2008,Datko2018}. The twelve animated games are designed with an overarching storyline that features the twelve Zodiac Animals. As Figure 1 (Middle) shows, the 12 zodiac animals live in the different locations in the game world. Each animal has a task that the player can complete by playing an NFT game. After helping all the animals, the player will unlock the final game, in which the Nian Beast (a mythological beast representing the end of the year) needs to be defeated with help from all the zodiac animals.

In each game, the player controls a team of characters, including a boy, a girl, and a sloth, to tackle a collaborative task in order to help a zodiac animal. For example, in the game with the mouse (Figure 3), the team needs to row a boat together to rescue the mouse who fell into the water by accident. With the neurofeedback mechanism, the player controls the team and aids the rescue by maintaining attention to the collaboration of the team. The decision for creating stories about different zodiac animals was supported by special educators’ observation and experience. Namely, they considered it an extra benefit if they were able to help the children memorize different animals while facilitating the training sessions. This has been proven in actual training: children showed ability to differentiate these animals after certain times of playing. All 12 games are designed with different collaborative tasks in a variety of scenes: for example, playing a ball passing game with the tiger (Figure 3), climbing the tree to save the monkey, or cheering on the snake in a dance battle against a fox (Figure 5). These collaborative tasks were also supported by prior research and our domain experts’ knowledge. Attending to interpersonal collaboration has been proven to be effective stimuli for autistic children’s mirror neuron activation \cite{Datko2018,Pineda2008,Direito2021}.

\begin{figure}[h]
  \centering
  \includegraphics[width=\linewidth]{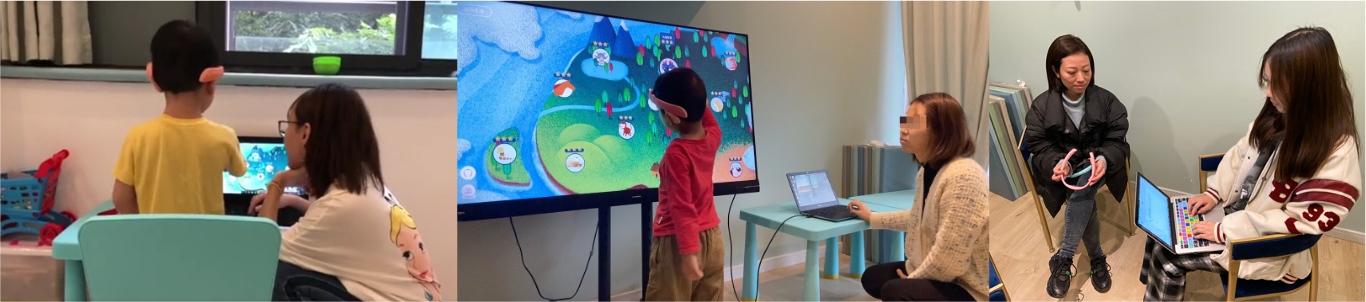}
  \caption{Left and Middle: caregivers facilitating the children in training sessions. Right: an interview with a child's parent}
  \Description{A woman and a girl in white dresses sit in an open car.}
\end{figure}

\section{Interdisciplinary collaboration and stakeholder involvement}

An interdisciplinary team has carried out a close collaboration to design and develop the Starkids system iteratively. The core project team consists of one neuroscientist, two algorithmic engineers (for signal processing and machine learning), two developers (for front-end and back-end development), and three designers (for interaction design and product design). In addition, a team of hardware engineers has been shared with other internal projects (for adapting the EEG headband based on product design requirements). A designer also serves the role of Product Manager to coordinate the team. 

\begin{figure}[h]
  \centering
  \includegraphics[width=\linewidth]{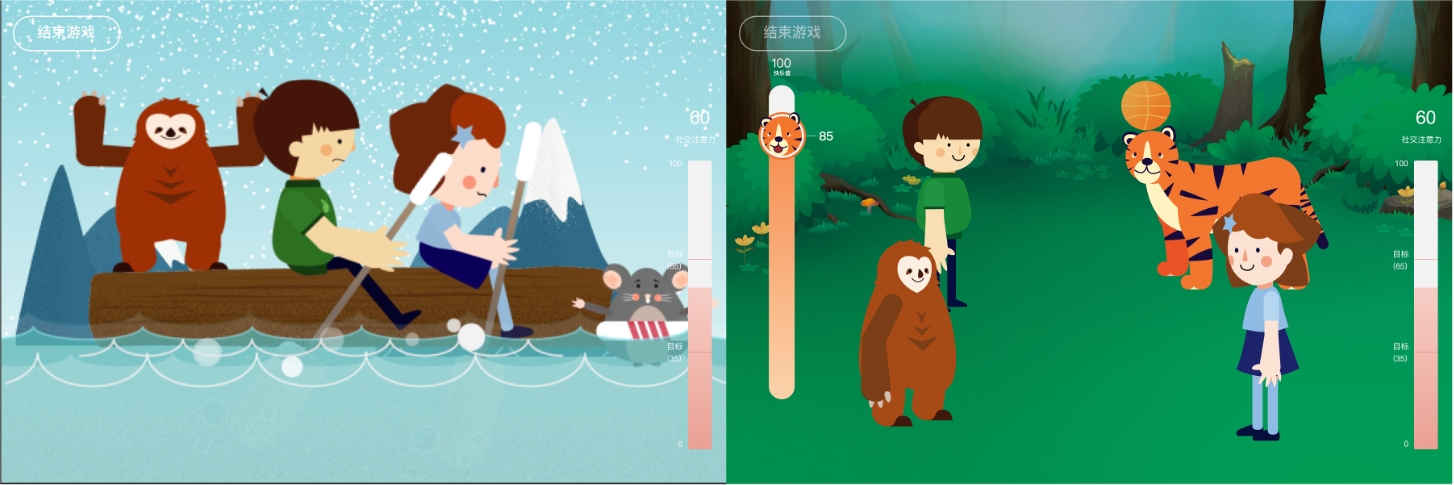}
  \caption{Left: ‘Rescuing the mouse’: the characters rowing a boat to rescue a mouse that fell into the water. Right: ‘playing with the tiger’: the characters play the ball-passing game together with a tiger.}
  \Description{Left: a boy, a girl and a slot rowing a boat to rescue a mouse who fell into the water; Right: a boy, a girl and a sloth playing ball passing game with a tiger.}
\end{figure}

Moreover, a large group of stakeholders (i.e., preschool children with ASD, their parents, and therapists) has been involved in the iterative development which partially took place in a special education center in Hangzhou, China. Nineteen children with ASD have been involved with their parents’ consent to use the Beta versions of Starkids in their existing training programs on a regular (weekly) basis. The period of each child’s involvement varied from two months to one year. The EEG data of the children interacting with the product has been gathered for function validation and therapeutic analysis (another ad-hoc clinical trial involving 60 children is ongoing). Twelve caregivers of the children, specialized in special education and applied psychology, were also involved regularly to provide feedback via interviews and focus groups (e.g., their observations and interpretations about the children’s reactions during the NFT games), or arrange field observations for the design team (Figure 2). On top of that, the design team also conducted a series of interviews with parents of the children, to gather their opinions and observations on how the system was used, as Figure 2 shows.

Above activities involving the stakeholders have brought great benefits to us, for establishing informative and efficient feedback loops between the real-world use contexts and the development. For this reason, quick tests and rapid iterative modifications of design features at detailed levels were enabled, and a set of design implications have been learnt from the data and insights gathered in the field (which are to be addressed in the next section). Apart from this, the knowledge of the caregivers was also effectively leveraged to stimulate our early design phase. Namely, their understandings about characteristics of children with ASD have helped us to generate the preliminary version of typical user profiles and design requirements, which directed and inspired the brainstorming of game ideas.

Reflecting on our process, we learned that the close involvement of caregivers and parents served as a key to comprehensively understand the experiences of the children, in addition to engaging the children directly in the field observations. There are several reasons for this. First, designing user experiences for preschool children in general is challenging, since they might not be able to accurately communicate their wants and desires \cite{Guha2010}. Yet designing for preschool children with ASD is even more challenging. On the one hand, there is much higher barrier for communication with these children (e.g., it can be rather challenging for many of them to retain concentration for a short interview, and certain questions can be too complex for them); and since designers are not qualified caregivers, they have to be extra careful in the communication in order to avoid unintentionally burdening the children. On the other hand, parents and caregivers have established deep understandings about the personality and preferences of the children over time, and thus are able to offer more accurate, and comprehensive interpretations about the children’s behaviors and experiences. Combining our field observation with their interpretations often led to valuable insights into how to iterate certain design features.

\section{Design Iterations and implications}

\subsection{Implication 1: emphasizing hand movements and simplifying facial expressions for characters}

As mentioned, the field observation and stakeholder feedback have inspired some of the game design features; One example of stakeholders’ valuable input to our iterative process concerns the facial features of the characters. The caregivers observed in practice that the children did not respond well to more intricate facial expressions, which they might feel uncomfortable or overwhelming. Moreover, our scientist pointed out that kids with ASD often avoid making eye-contacts; forced eye-contacts can cause elevated anxiety for them. It was therefore suggested that the characters should be given very simplistic facial features. We applied this insight by giving the characters two dots as eyes and a simple curve or shape for a mouth, the minimal viable features needed to recognize a face (see Figure 4). In this way, the children showed willingness to engage with the characters with little avoidance or anxiety.

 Another example is the iteration of the hands of the cartoon characters in the games. When designing characters for the games, we were advised by caregivers to make sure that actions performed by the characters were easy to be noticed and understood. Based on their observation, a number of the children seemed to show extra interests in the hand activities of the characters in NFT games. The neuroscientist also shared that the mirror neuron activity of an autistic child could be easily triggered when the child was watching others’ hand movement such as hands using tools (e.g., see an experiment design in \cite{Datko2018}). As Figure 4 shows, we thereby gave the characters larger hands and more emphasized hand movement, to facilitate engagement and active response from the children. While these observations and knowledge-in-practice from domain experts have greatly benefited us in establishing such general principles for iterations, a future step for us is to create nuanced comparisons to continue extracting more elaborate design guidelines (e.g., the sweet spots of configuring facial elements or hand movement of characters).

\begin{figure}[h]
  \centering
  \includegraphics[width=\linewidth]{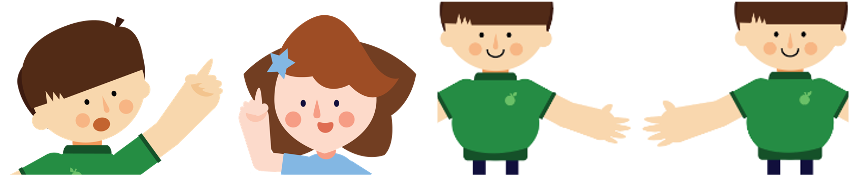}
  \caption{Left: the simplistic design of expressions and eye behaviors. Right: enhanced hand proportion.}
  \Description{Left: the detailed look of the facial expressions designed for the main characters; Right: the hands of the characters have been enlarged after a design iteration.}
\end{figure}

\subsection{Implication 2: diversifying gameplay experiences while having a coherent overarching story}

This is a relatively higher-level design implication we generalized to help NFT game systems to better maintain engaging experience for children with ASD, both in the short-period (each training session) and over time (during a training program). Essentially, all NFT games for children with ASD follow the same basic mechanism, in which the player gets rewarding feedback by retaining social attention (mirror neuron activation). A challenge for designers, thereby, is how to create attractive player experience with this simple mechanism, to continually engage the children in a training session, or across sessions over time. 

\begin{figure}[h]
  \centering
  \includegraphics[width=\linewidth]{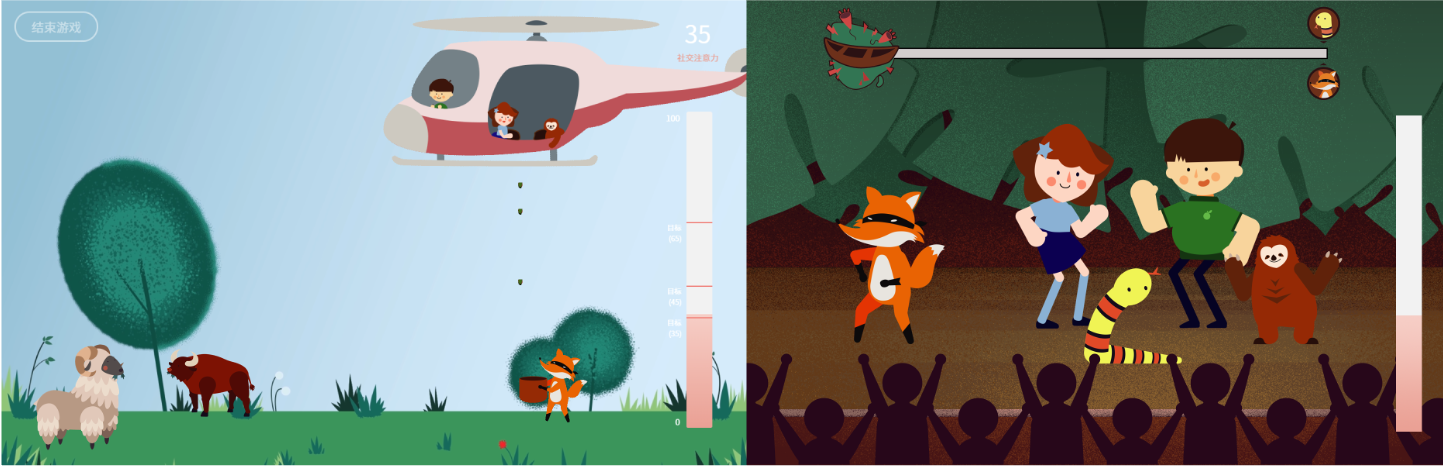}
  \caption{Left: ‘Helping the ox and the sheep’; Right: ‘Snake versus fox dance-battle’.}
  \Description{Left: the main characters are flying a helicopter to help the ox and the sheep on the ground; Right: the main characters dancing together with a snake to battle a fox.}
\end{figure}

Based on our explorations with the stakeholders (the children, their caregivers, and parents) in real-world training programs, we summarize two key points: (1) diversifying gameplay experiences by providing a series of games that differ in gameplay designs and have varying scenes; (2) connecting the series of games with a coherent overarching story (consistent main characters and background). These two points were also embodied in our design of the ``Zodiac Animals'' game series (Figure 1), which contains 12 different games and scenes, under a simple and coherent setting (three main characters helping each animal out of its own trouble by completing the first eleven games and beat the Nian beast together at the final game). Based on our field observations with caregivers, the benefit of such design strategy is twofold. First, the variety of gameplay designs and scenes gave the children rich choices each time and avoided them feeling bored of only playing one game again and again. Second, the unified overarching story (background) helped engage the children in the long run, since it enabled them to establish personal connection with the same group of characters over time, across a coherent series of quests (helping different animals). As observed, the children were able to memorize the characters after certain time, and responded with interest when their facilitator mentioned the characters.

Our observations could also be supported by related research focusing on an older group of children with ASD (8-13 years old) \cite{Mercado2020}, which reported that continue playing a game may cause the children to show boredom towards the end of a session, despite the game being rather enjoyable in the beginning. Beyond that, our caregivers also observed that some children tended to repetitively play the same game that they were accustomed to, which could reflect the repetitive, rigid behavior patterns of ASD. From a caring perspective, such repetition would not need to be reinforced, and this could be another aspect that is to be resolved by creating diverse gameplay options for these children. Nonetheless, related research so far has mostly focused on short-term user experiences of a single NFT game (designed for research purposes instead of as a product). We hence argue that our kind of design research is also needed, to understand the children’s long-term player experiences of a series of games within a real-world service system in existing special education settings.

\subsection{Implication 3: streamlining game narrative design for autistic children}

For preschool children with ASD, the simplicity of the game narratives is as important as the diversity in gameplay. Through observations of how the children engaged with different games designed by us, we have also accumulated practical understandings about how simplistic the game narratives should be to balance the learnability and richness of the game. Autistic children with cognitive impairment can have difficulty in dealing with complex game concepts. Simple, short, and straightforward narratives can help them understand the rules of the games (as also supported by \cite{Mandryk2013}). Meanwhile, they can be easily distracted by excessive game elements due to often co-occurring attention deficits. Taken together, an important implication is to maintain an appropriate level of simplicity in the game narrative design, so that they could easily understand the core story and mechanism, and have minimal distraction from unnecessary elements.

For example, the game of ‘Helping the ox and the sheep’ (Figure 5) asks players to aid the ox and the sheep by retaining certain level of social attention (indicated by mu waves) to the characters. When the social attention index is lower than the preset threshold, there will be a fox to interfere with the mission. During the observations, it was found that some children had difficulty understanding the idea of the game and they were easily distracted by the fox. The fox was thereby removed to help the users better grasp the core story. By contrast, the game ‘Rescuing the mouse’ (Figure 3) has very simple and straightforward mechanism: to rescue the mouse in the water, a user needs to maintain social attention to the main characters rowing the boat. With higher level of user’s social attention, the characters row faster to complete the rescue. As observed, users could understand and complete this task very well.

Another example of simplifying the game concept for preschool children could be reflected by our strategy of designing rewards. Rewards are in general highly important in gamification. However, our explorations showed that some traditional approaches of rewards (e.g., scores, coin counts, treasure chests, or badges) may not be the most suitable real-time rewards to our target children. First, these approaches often introduce new concept in the gameplay, which adds to the complexity for preschool autistic children to comprehend. Second, these approaches also introduce new visual elements (e.g., a number for counting the points/coins collected in game), which can be distractions during the game for these children (from observation we found that numbers can be especially attractive to children with ASD; however, attending to numbers will not help mirror neuron activation). Therefore, in our case, we explored designing the expressive moves of main characters as the instant rewards for players’ attention reaching different levels. For instance, in the “Snake versus fox dance-battle” game, the snake (as central character) will play a somersault once the player’s mu wave index has reached a certain level and kept it for a few seconds. As the observations showed, this was not difficult for the children to achieve, and they were very excited to see the expressive move of the snake (some children jumped together with the snake), and thereby motivated to continue engaging with the characters. This way, it created a positive feedback loop for their concentration. Similar examples were observed by us in other games as well, which indicated that designing the central characters’ expressive moves as instant rewards to the children’s attention could be a good strategy to leverage the benefits of rewards (e.g., motivating engagement) without distracting their focus during a NFT game.

\subsection{Implication 4: leveraging multisensorial ensemble for neurofeedback games}

From our design explorations, we found that complementing visual feedback with auditory feedback is a good design strategy. Auditory feedback could synchronize with the visual feedback to increase the immersion of players or the salience of their reward. In addition, auditory feedback could replace certain visual feedback to reduce the information load of visual channel, since children with ASD could be easily distracted from the intended focus of the gameplay by excessive visual elements on the screen (e.g., the fox in ‘Helping the ox and the sheep’, see Figure 5). 

An example of such meaningful combination can be found in the game of ‘Building a house for the little pig’, in which there are many different sound effects: such as tapping wood or tiles. We extracted real-world sounds to increase the immersion. The rhythm and types of sound effects change corresponding to the player’s social attention level. The ‘Snake versus fox dance-battle’ (see Figure 5) is another example. We map three different sound effects to three different dance moves determined by players’ social attentional levels. Namely, the higher the social attention score players get, the fancier a dance move appears along with the stronger rhythm of the music. Hence, users receive positive feedback synchronizing both visual and auditory stimuli. This also inspires us to integrate more multisensorial experiences in the future, such as haptic feedback.

\section{Conclusion and future work}

In this work we briefly reviewed our practice experiences gained from the Starkids (alias) project. Specifically, we focused on some concrete design implications on how to create enjoyable and playful experiences for NFT games targeted at children with ASD in special education contexts. Two major takeaway insights might benefit the future practice in the application domain. The first insight we learnt is the importance of engaging both the target children and the associated stakeholders (the children’s caregivers and parents). We found that to comprehensively understand the experiences of children with ASD, in addition to direct engagement, it is also necessary to have in-depth involvement of their caregivers and parents. This is not only due to the challenge of communication, but also the significant individual differences existing among the children. Such individual differences and background stories can be the key to interpret certain responses and preferences of each child, which could be greatly facilitated by the long-term understandings established by the children’s caregivers and parents. The second insight we hope to stress is the importance of understanding the differences between designing NFT games for children with ASD and designing edutainment games in general. For instance, as we found, some gamification elements that are commonly seen in general edutainment games may not be as effective in NFT games for children with ASD. As mentioned, a typical rewarding element, such as a number for counting score or coins earned, might distract the autistic children’s attention away from the intended focus of the gameplay (the collaboration of the characters). Moreover, as we found, the rewarding mechanism should be designed to be simple enough for the target children to understand (considering that some low-functioning autistic children may have more difficulty in understanding complex stories/concepts). In addition, we observed that autistic children are especially interested in rewarding feedback that is explicit and immediate; by contrast, they are less responsive to feedback that is more nuanced or gradual. These examples give a glimpse about how the design of NFT games for autistic children might differ from the design of general edutainment games.

While we hope these implications could advance the design of such systems, we also note that the scope of such a product-service system is beyond designing the NFT games per se. Namely, in the ongoing collaboration with the special education center, we are also learning how to safely leverage the training data to improve the care and education quality, e.g., by offering meaningful analytics tools (Figure 6) to aid the insights of caregivers, program managers, or parents. Moreover, a parallel exploration on optimizing the hardware design is also in progress (Figure 6). We believe in the near future, we will be able to provide a more comprehensive perspective to address relevant implications on what it means to embed a neurofeedback product-service solution into current special education contexts.

\begin{figure}[h]
  \centering
  \includegraphics[width=\linewidth]{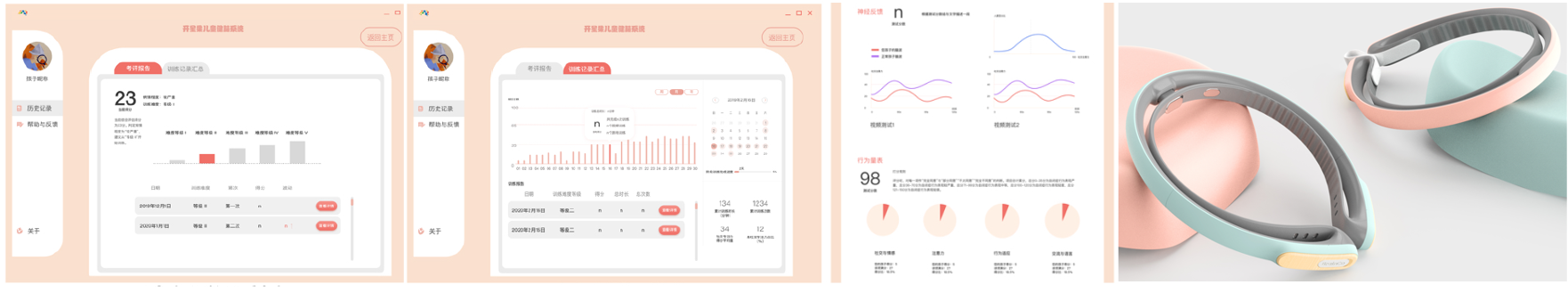}
  \caption{Left Three: analytic dashboard screens for caregivers. Right: new headset design with increased wearability and comfort.}
  \Description{Left three images show the dashboard interfaces designed for the caregivers to monitor and anlyze the training progress of the children with ASD. Right image shows two newly designed headsets with increased comfort and wearability.}
\end{figure}

\begin{acks}
We thank our colleagues from BrainCo and our collaborators from the special education center, and all the children, caregivers and parents who involved in this work.
\end{acks}

\bibliographystyle{ACM-Reference-Format}
\bibliography{references}

\end{document}